\documentclass[pra]{revtex4}
\usepackage{epsfig,graphicx,tabularx}
\usepackage{graphicx}
\usepackage{graphics}
\usepackage{epsfig}
\usepackage{amsmath,amssymb,mathrsfs}
\usepackage{amsmath}
\usepackage{amsfonts}
\usepackage{amssymb}
\usepackage{verbatim,xcolor,ulem}

\newcommand{\beq}{\begin{equation}}
\newcommand{\eeq}{\end{equation}}
\newcommand{\beqa}{\begin{eqnarray}}
\newcommand{\eeqa}{\end{eqnarray}}
\newcommand{\ba}{\begin{aligned}[b]}
\newcommand{\ea}{\end{aligned}}

\newcommand{\Ce}{{\rm C}}
\newcommand{\GCe}{{\rm GC}}

\newcommand{\be}{\begin{equation}}
\newcommand{\ee}{\end{equation}}
\newcommand{\bes}{\begin{equation*}}
\newcommand{\ees}{\end{equation*}}

\newcommand{\pder}[2]{ { \frac{\partial}{\partial #1}\Bigr|_{#2} }}

\newcommand{\ov}[1]{\overline{#1}}
\newcommand{\eps}{\epsilon}
\newcommand{\E}{\mathrm{e}}

\begin{document}

\title{Canonical vs. Grand Canonical Ensemble \\ for Bosonic
  Gases under Harmonic Confinement}

\author{Andrea Crisanti$^{1}$, Luca Salasnich$^{2,3,4,}$,
  Alessandro Sarracino$^{5}$ and Marco Zannetti$^{6}$}

\affiliation{$^{1}$Dipartimento di Fisica, Universit\`a di
  Roma ``La Sapienza'', Piazzale Moro 5, 00185 Roma, Italy \\
  $^{2}$Dipartimento di Fisica e Astronomia ``Galileo Galilei''
  and Padua QTech Center, Universit\`a di Padova, Via Marzolo 8, 35131
  Padova, Italy \\
  $^{3}$Istituto Nazionale di Fisica Nucleare, Sezione di Padova,
  Via Marzolo 8, 35131 Padova, Italy\\
  $^{4}$Istituto Nazionale di Ottica del Consiglio Nazionale
  Delle Ricerche, Via Nello Carrara 2, 50127 Sesto Fiorentino, Italy\\
  $^{5}$Dipartimento di Ingegneria, Universit\`a della Campania
  ``Luigi Vanvitelli'', Via Roma 29, 81031, Italy\\
  $^{6}$Dipartimento di Fisica ``Eduardo Caianiello'',
  Universit\`a di Salerno, Via Giovanni Paolo II 132, 84084 Salerno}

\begin{abstract}
  We analyze the general relation between canonical and grand canonical ensembles in the thermodynamic limit. We begin our discussion by deriving, with an alternative approach, some standard results first obtained by Kac and coworkers in the late 1970s. Then, motivated by the Bose--Einstein condensation (BEC) of trapped gases with a fixed number of atoms, which is well described by the canonical ensemble and by the recent {groundbreaking} experimental realization of BEC with photons in a dye-filled optical microcavity under genuine grand canonical conditions, we apply our formalism to a system of non-interacting Bose particles confined in a two-dimensional harmonic trap. We discuss in detail the mathematical origin of the inequivalence of ensembles observed in the condensed phase, giving place to the so-called grand canonical catastrophe of density fluctuations. We also provide explicit analytical expressions for the internal energy and specific heat and compare them with available experimental data. For these quantities, we show the equivalence of ensembles in the thermodynamic limit.
  \end{abstract}

\maketitle

\section{Introduction}

Bose--Einstein condensation (BEC) was first experimentally observed with ultracold and dilute atomic gases in 1995 by several groups~\cite{bose1,bose2,bose3}. Since then, the phenomenon has been reproduced in a huge variety of systems~\cite{cornell2002nobel}, such as trapped alkali gases~\cite{leggett2001bose,dalfovo1999theory}, rotating systems~\cite{fetter2009rotating}, quantum magnets~\cite{zapf2014bose}, and so on. For all these systems,
the number of atoms is conserved and the statistical properties are well described within the canonical ensemble (CE).  
More recently, however, and quite surprisingly, BEC has also been realized with effectively massive photons in a dye-filled optical cavity \cite{klaers2010bose,klaers2011bose,schmitt2014observation,damm2016calorimetry,damm2017first,schmitt2018dynamics,ozturk2023fluctuation}, where it is possible to work in a regime with a number of photons that are conserved only on average. In this case, the proper ensemble to describe the system features is the grand canonical ensemble (GCE). 

The relation between CE and GCE is a general problem of statistical mechanics and the two ensembles are usually equivalent in the thermodynamic limit. This is the case when 
correlations are short-ranged and, due to the central limit theorem,
fluctuations in extensive quantities become
negligible for a large system size. 
However, there exist noteworthy exceptions, where computations performed in the two ensembles lead to different results. This occurs, for instance, in long-range systems~\cite{campa2009statistical}, where the issue of ensemble inequivalence is very well studied. In the framework of systems of free bosons, this topic has been considered theoretically in several studies~\cite{ziff1977ideal,holthaus1998condensate,fujiwara1970fluctuations,kocharovsky2006fluctuations,yukalov2007bose}. The remarkable result is that, while, in the CE, density fluctuations vanish in the thermodynamic limit as generally expected, in the GCE, they are macroscopic and remain finite upon decreasing the temperature, leading to the so-called grand canonical catastrophe, with an explicit negative connotation. However, the experimental realization of BEC in genuine grand canonical conditions has confirmed the theoretical predictions obtained in the GCE, making the issue related to the inequivalence of ensembles in this context a heated topic of debate. In a previous paper~\cite{crisanti2019condensation},  building on an analogy with the celebrated spherical model of a ferromagnet introduced by Berlin and Kac~\cite{berlin1952spherical}, some of us have interpreted the physical meaning of the grand canonical catastrophe predicted for a homogeneous ideal Bose gas as a phenomenon of the condensation of fluctuations~\cite{zannetti2015grand}.

In this paper, before considering the experimentally relevant case of quantum gases of non-interacting bosons confined 
in a two-dimensional harmonic trap, we first present an accurate discussion of
the relation between canonical and grand canonical ensembles in the general framework of statistical mechanics. In particular, we bring to the fore the mathematical aspects related to the (in-)equivalence between ensembles, focusing on the behaviors arising in the thermodynamic limit. We re-derive some important results obtained by Ziff et al.~\cite{ziff1977ideal} for ideal gases, following an alternative approach, based on the theory of singular perturbations and boundary layer problems~\cite{holmes2012introduction}. We then apply our formalism to the case of a free gas of bosons confined by a parabolic potential and investigate several physical features. 
We find that, in the thermodynamic limit, in the Bose-condensed phase, 
the density fluctuations and the spatial density--density correlation function 
have a quite different behavior in the two ensembles. In contrast, other quantities, 
such as the critical temperature, the condensate fraction, internal energy, and the specific heat behave similarly. For the last two quantities, we provide explicit analytical expressions that very well compare to the experimental data reported in~\cite{damm2016calorimetry}. 

This paper is structured as follows: in Section \ref{Sec:Gen}, we develop the general formalism to connect CE to GCE 
with the kernel introduced in ref. \cite{kac1977correlation}. In Section 
\ref{Sec:Ideal}, we discuss the problem of Bose--Einstein condensation for an ideal gas of massive bosons 
confined in a two-dimensional harmonic potential. In Section \ref{Sec:Ideal}, we compare the theoretical predictions of the CE and GCE for several quantities. {We discuss in particular the issue related to the grand canonical catastrophe and its meaning.} 
We also show that our theory reproduces the experimental results of internal energy and specific heat 
of the photons in {an} optical cavity quite well. Conclusions are drawn in Section \ref{Sec:Conc}. It is important to stress that, in our two-dimensional problem, there is true BEC due to the presence of the harmonic confinement, while there is no  Berezinskii--Kosterlitz--Thouless phase transition because the bosons  are not interacting.

\section{Relation between Canonical and Grand Canonical Ensembles}
\label{Sec:Gen}

{We start by discussing some standard results of statistical mechanics. This allows us to introduce the general problem and to fix the notation that will be used throughout the paper.}
A general relation between the CE and GCE can
be established at different levels. We start by considering the
connection between the partition functions. We 
denote by ${\sigma}=(\sigma_1,\ldots,\sigma_N)$ the state of a generic system, and by ${\cal H}({\sigma})$ its Hamiltonian.
The natural variables of the CE are the temperature $k_B T =\beta^{-1}$, the volume $V$, and 
the number of elements $N$. It is important to underline that, in the subsequent sections, which investigate the bosonic system under harmonic confinement, the volume $V$ will be an ``effective volume'' related to the frequency of the harmonic potential. The probability density of a configuration ${\sigma}$ is 
\be
\label{eq:rhoC}
  P_\Ce({\sigma}|\beta,V,N)  = \frac{1}{Z_\Ce(\beta,V,N)}\, \E^{-\beta {\cal H}({\sigma})},
\ee
where
\be
  Z_\Ce(\beta,V,N) = \sum_{\{{\sigma}\}}\, \E^{-\beta {\cal H}({\sigma})}
\ee
is the canonical partition function, where
the sum extends over all possible system configurations.

In the GCE, the natural variables are the temperature 
$k_B T=\beta^{-1}$, the volume $V$, and the fugacity $z=e^{\beta\mu}$,
where $\mu$ is the chemical potential.
The probability that the system is in the configuration ${\sigma}$ with $N$ elements
is, in the GCE,
\be
\label{eq:rhoGC}
P_\GCe({\sigma},N|\beta,V,z)  = \frac{z^N \E^{-\beta {\cal H}_{N}({\sigma})}}{Z_\GCe(\beta,V,z)},  
 \ee
 where
 \beqa
\label{eq:ZGC}
Z_\GCe(\beta,V,z) &=& \sum_{N\geq 0}\, \sum_{\{{\sigma}\}}\, z^N\, \E^{-\beta {\cal H}_{N}({\sigma})} \nonumber \\
&=& \sum_{N\geq 0}\, z^N Z_\Ce(\beta,V,N)
\eeqa
  is the grand canonical
  partition function defined for $\Re z < z_0$, where $z_0$ is a model-dependent parameter ensuring the convergence of the sum in Equation~(\ref{eq:ZGC}).
We have added the subscript $N$ to the Hamiltonian to indicate that 
${\cal H}_{N}({\sigma})$ refers to  a system of $N$ elements.
  From the above relation, one immediately has
  that the GCE partition function can be obtained as a $z-$transform of
  the CE partition function.  The Relation \eqref{eq:ZGC} can be
  inverted~as 
\be
\label{eq:ZCe_ZGCe}
Z_\Ce(N) = \frac{1}{2\pi i} \oint_{\Gamma} dz \, z^{-1-N}\, Z_\GCe(z),
\ee where $\Gamma$ is a generic curve in the complex plane encircling
the origin $z=0$ (to simplify the notation, we have only indicated the
relevant variables).

\subsection{The Kernel $K(N|z)$}

The connection between the GCE and CE can also be established at the level
of probability density. Indeed, writing
\be
  P_\GCe({\sigma},N|z)  = \frac{z^N\, Z_\Ce(N)}{Z_\GCe(z)}\,   
                    \frac{\E^{-\beta {\cal H}({\sigma})}}{Z_\Ce(N)},
\ee
and using \eqref{eq:rhoC}, one obtains
\be
\label{eq:rhoGC-rhoC}
   P_\GCe({\sigma},N|z) =  P_\Ce({\sigma}|N)\, K(N|z),
\ee
where
\be
\label{eq:KNz}
  K(N|z) = \frac{z^N\, Z_\Ce(N)}{Z_\GCe(z)}
\ee
is the kernel
  relating the two distributions, and represents the probability that the system consists of $N$ elements for fixed $z$. Its explicit form depends on the
  specific model. Celebrated examples are discussed in~\cite{kac1977correlation} for the spherical model and in~\cite{ziff1977ideal} for the homogeneous ideal Bose gas. We will consider the trapped ideal Bose gas in
  the following sections.

Finally, from the Relation
  \eqref{eq:rhoGC-rhoC}, a connection between averages readily follows. Indeed, we have
  \be 
  \label{eq:ave}
  \langle
  f({\sigma})\rangle_\GCe(z) = \sum_{N} \langle
  f({\sigma})\rangle_\Ce(N)\, K(N|z), \ee where we have denoted by $
  \langle \ldots\rangle_\GCe$ and $\langle \ldots\rangle_\Ce$ averages
  in the GCE and CE, respectively, and $f({\sigma})$ is a generic
  function of the system configuration. {\color{black}The above equation can also be interpreted as the generating function of $\langle
  f({\sigma})\rangle_\Ce$, as first introduced in~\cite{ziff1977ideal}; see their Equation~{(2.34})}.
  As detailed in Appendix A,
  this relation can be formally inverted, yielding
\be
\langle f({\sigma})\rangle_\Ce(N) =  \frac{1}{2\pi i}\, \oint_{\Gamma} \frac{dz}{z} 
                    \frac{\langle f({\sigma})\rangle_\GCe(z)}{K(N|z)}.
                    \ee

\subsection{Thermodynamic Limit}
\label{sec:Th_Lim}
 
The discussion presented in  the previous section can be simplified in the large $N$ limit,
leading to some more explicit general results. 
The thermodynamic limit is obtained by~taking
\be
\label{eq:ThLim}
  N\gg1,\ V\gg 1\ \text{but}\ N/V=\rho = \text{constant}.
\ee

Rewriting the GCE partition function \eqref{eq:ZGC} as
\be
\label{eq:ZGC_1}
  Z_\GCe(z,V) = \sum_{N} \E^{N\ln z + \ln Z_\Ce(N,V)},
\ee
since $\ln Z_\Ce(N,V) = O(N)$ as $N\gg1$,   
the sum is dominated in the thermodynamic  limit~\eqref{eq:ThLim} by the term with the largest value of the  
exponent:
\be
\label{eq:ZGCTL1}
   Z_\GCe(z,V)  \sim z^{N^*(z)}\, Z_\Ce(N^*(z),V),
\ee
where
\be
N^*(z) = max_N[N\ln z + \ln Z_\Ce(N,V)\bigr].
\ee

To evaluate $N^*(z)$, we rewrite Equation \eqref{eq:ZGC_1} as
\be
   Z_\GCe(z,V) = V\, \sum_{N}\, \Delta\rho\, \E^{V\bigl[\rho\ln z + \frac{1}{V}\ln Z_\Ce(\rho,V)\bigr]},
\ee
where $\Delta \rho = 1/V \ll 1$ as $V\gg 1$, and, in the thermodynamic limit, we have
\be
   Z_\GCe(z,V) \sim V \int d\rho\, \E^{V \phi(\rho)},  \,\, V\gg1
\ee
with
\be
\label{eq:phir_1}
  \phi(\rho) = \rho\ln z + \frac{1}{V}\ln Z_\Ce(\rho,V).
  \ee
  
  Note that, since $\ln Z_\Ce(\rho,V)=O(V)$ in the large volume limit, the second term on the right-hand side becomes independent of $V$. We will drop $O(1/V)$ corrections in the following.
The integral can then be evaluated using the Laplace method, expanding $\phi(\rho)$ as
\be
\label{eq:phir_2}
  \phi(\rho) = \phi(\rho^*) + \frac{1}{2}\phi''(\rho^*)(\rho - \rho^*)^2 + O(\rho-\rho^*)^3,
\ee
where $\rho^*$ is the solution of the stationary point equation
\be
  \phi'(\rho^*) =  \ln z + \frac{1}{V}\frac{\partial}{\partial \rho^*}\ln Z_\Ce(\rho^*,V) = 0.
\ee

Then, introducing $\sigma_\phi^{-2} = -(\partial/\partial\rho^*)^2 \ln Z_\Ce(\rho^*,V) > 0$, we have
\be
  Z_\GCe(z,V) \sim   V\,\sqrt{2\pi \sigma_\phi^2}\, \E^{V \bigl[\rho^*\ln z + \frac{1}{V}\ln Z_\Ce(\rho^*,V)\bigr]},
   V \gg1.
\ee

Writing $N^*(z) = V\rho^*$, and taking only the leading term, we recover Equation~\eqref{eq:ZGCTL1}, 
with $N^*(z)$ given by the solution of
\be
\label{eq:Ce-GCe-Leg0}
  N^*(z):\ \frac{\partial}{\partial N} \ln Z_\Ce(N,V) = -\ln z.
\ee

Equation \eqref{eq:ZGCTL1} implies
that
\be
\label{eq:Ce-GCe-Leg1}
  \ln Z_\GCe(z,V) = \ln Z_\Ce(N,V) + N\ln z,
\ee
where $N=N^*(z)$ is obtained from Equation~\eqref{eq:Ce-GCe-Leg0}.
Equations \eqref{eq:Ce-GCe-Leg0} and \eqref{eq:Ce-GCe-Leg1} show that,
in the thermodynamic limit \eqref{eq:ThLim},  the $ \ln Z_\GCe(z,V)$ is the Legendre 
transform of $\ln Z_\Ce(N,V)$  with respect to $N$.

We now consider the kernel $K(N|z)$. From the previous discussion, it follows that, in the thermodynamic limit, for fixed $z$,
\beqa
\label{solo-lui}
   K(N|z) &=& \frac{z^N\, Z_\Ce(N,V)}{Z_\GCe(z,V)} 
              = \frac{\E^{V\phi(\rho)}}{V \int d\rho\, \E^{V\phi(\rho)}} \nonumber \\
              &\sim& \frac{\E^{-\frac{1}{2\sigma_\phi^2}(\rho - \rho^*)^2}}{V\sqrt{2\pi\sigma_\phi^2}},
\eeqa
where, in the second equality, we have used Equation \eqref{eq:phir_1},
and hence, since $\sigma^2_\phi = O(1/N)$, 
\be
\label{eq:KNz_dir}
   K(N|z) \sim \Delta\rho\, \delta(\rho -\rho^*) \sim \delta^{\rm Kr}\bigl(N-N^*(z)\bigr),
\ee
with $N^*(z)$ the solution of \eqref{eq:Ce-GCe-Leg0} 
and $\delta^{Kr}$ the Kronecker delta. This result is also discussed,
for instance, in~\cite{ziff1977ideal}---see their Equation~({2.63})---and shows the equivalence of the ensembles in the thermodynamic limit.
The same result can be obtained starting from the expression of $Z_\Ce(N)$ as a transform
of $Z_\GCe(z)$, as reported in Appendix B {of this paper}. The above discussion shows that, for a fixed $z$, with $\Re z < z_0$,
there exists a corresponding $N^*(z)$ and therefore there exists
an equivalent canonical ensemble with $N=N^*(z)$.

The general findings obtained so far break down in specific cases,
where the application of the Laplace method requires particular
care. This can lead to different behaviors in different ensembles, as
described in the next section.

\section{Ideal Bose Gas in a Harmonic Trap}
\label{Sec:Ideal}

As a specific system, we consider a two-dimensional (2D) harmonically
trapped system of non-interacting massive bosons. This model has been
shown to provide a very good description of a photon gas confined in a
high-finesse dye-filled microcavity, studied in the
experiments~\cite{klaers2010bose,klaers2011bose,schmitt2014observation,damm2016calorimetry,damm2017first,schmitt2018dynamics,ozturk2023fluctuation}, where Bose--Einstein
condensation has been observed. In particular, the microcavity was
realized by two curved mirrors, where photons were continuously
absorbed and re-emitted by the dye medium, {which is crucial for the equilibration}. Due to a cavity length of
the same order as the photon wavelength, a frequency gap between the
longitudinal resonator modes was realized. Thus, the system consists
of photons with a fixed longitudinal mode number that are allowed to
populate several transversally excited cavity modes, making the photon
gas effectively two-dimensional. The energy--momentum relation is
described by a quadratic term, with an effective mass, and a trapping
parabolic potential induced by the mirror curvature~\cite{klaers2011bose}.

It has been shown that this experimental setup realizes bona fide grand canonical
conditions where theoretical predictions can be tested \cite{damm2017first,schmitt2018dynamics,ozturk2023fluctuation}. 
We are interested in the different behaviors that can arise in
different ensembles, in particular for what concerns the fluctuations
in the occupation number of the condensed phase. In the following, we
will apply the formalism introduced in the previous section to the
system of trapped bosons in 2D. Our theoretical treatment differs from
that presented in~\cite{klaers2010bose}, where the photon number
statistics are derived from a rate equation model in the stationary
regime. Our analysis follows rather the analytical theory of Ziff et
al. \cite{ziff1977ideal}, developed for the case of a free ideal
gas. The main difference relies on the derivation of the Kac kernel in
the region below the critical temperature, which we present here for the
case of a trapped Bose gas, {\color{black} following an alternative approach based on the theory of singular perturbations}.

\subsection{The Model}

In second quantization, the Hamiltonian of the system can be 
written {as} \cite{salasnich2017quantum}
\beq 
{\hat H} = \int d^2{\bf r} \ {\hat \psi}^{\dagger}({\bf r}) 
\left[ -{\hbar^2\over 2m_\textrm{eff}} \nabla^2 + U({\bf r}) 
\right] {\hat \psi}({\bf r}) \; , 
\label{hamilton}
\eeq 
where ${\hat \psi}({\bf r})$ (${\hat \psi}^{\dagger}({\bf r})$) is the bosonic field operator,  
which destroys (creates) a boson at the position ${\bf r}=(x,y)$, 
$m_\textrm{eff}$ is the effective mass of each photon, which depends on the mirror separation and on the linear index of the refraction of the medium (see~\cite{klaers2011bose} for details),
and 
\beq 
U({\bf r}) = {1\over 2} m_\textrm{eff} \omega^2 (x^2 + y^2) 
\label{harm}
\eeq
is the 2D harmonic potential with trapping frequency $\omega$. 
The local number density operator is given by 
\beq 
{\hat \rho}({\bf r}) = {\hat \psi}^{\dagger}({\bf r}) {\hat \psi}({\bf r}),
\eeq
while 
\beq 
{\hat N} = \int d^2{\bf r} \, {\hat \rho}({\bf r}) 
\eeq
is the total number operator. 

The single-particle quantum mechanics in the 2D harmonic potential 
(\ref{harm}) are described by the stationary Schr\"odinger equation 
\beq 
\left[ -{\hbar^2\over 2m_\textrm{eff}} \nabla^2 + U({\bf r})  \right] 
\phi_{\bf m}({\bf r}) = \epsilon_{\bf m} \, \phi_{\bf m}({\bf r}) \; , 
\label{sse}
\eeq
where ${\bf m} = (m_x,m_y)$ and the eigenvalues are given by 
$\epsilon_{\bf m} = \hbar \omega (m_x + m_y + 1)$
 with $m_x,m_y=0,1,2,...$ being the two natural quantum numbers, while 
$\phi_{\bf m}({\bf r})$ are the corresponding orthonormal eigenfunctions. 
In particular, for the single-particle ground state, we have 
$\epsilon_{\bf 0} = \hbar \omega$
and 
\beq 
\phi_{\bf 0}({\bf r}) = {1\over \sqrt{\pi} l_H} e^{-{x^2+y^2\over 2 l_H^2}},
\label{condwave}
\eeq
with 
$l_H = \sqrt{\hbar\over m_\textrm{eff} \omega}$
the characteristic length of the harmonic confinement. 

The field operator ${\hat \psi}({\bf r})$ can be expanded in any orthonormal 
basis. Here, we choose the basis of the eigenfunctions 
$\phi_{\bf m}({\bf r})$ of the single-particle problem of Equation (\ref{sse}), 
i.e., 
\beqa 
{\hat \psi}({\bf r}) &=& \sum_{{\bf m}\in \mathbb{N}^2}  
{\hat a}_{\bf m} \phi_{\bf m}({\bf r}), 
\label{expand1}
\\
{\hat \psi}^{\dagger}({\bf r}) &=& \sum_{{\bf m}\in \mathbb{N}^2}  
{\hat a}_{\bf m}^{\dagger} \phi_{\bf m}^\ast({\bf r}), 
\label{expand2}
\eeqa
where ${\hat a}_{\bf m}$ and ${\hat a}_{\bf m}^{\dagger}$ are the ladder operators 
which, respectively, destroy and create a boson in the single-particle 
quantum state $\phi_{\bf m}({\bf r})$. Inserting the Formulas (\ref{expand1}) 
and (\ref{expand2}) into Equation~(\ref{hamilton}), and using Equation (\ref{sse}) and the orthonormal property, we obtain 
\beq 
{\hat H} = \sum_{{\bf m}\in \mathbb{N}^2}
\epsilon_{\bf m} \ {\hat N}_{\bf m} \; , 
\label{hamsimple}
\eeq
where 
\beq 
{\hat N}_{\bf m} = {\hat a}_{\bf m}^{\dagger} {\hat a}_{\bf m}
\eeq
is the number operator of the single-particle 
quantum state. Similarly, we find 
\beq 
{\hat \rho}({\bf r}) = \sum_{{\bf m},{\bf m}^\prime \in \mathbb{N}^2}
{\hat a}_{\bf m}^{\dagger} {\hat a}_{{\bf m}^\prime} \phi_{\bf m}^\ast({\bf r}) \phi_{{\bf m}^\prime}({\bf r}),
\eeq
and clearly 
\beq 
{\hat N} = \sum_{{\bf m}\in \mathbb{N}^2}
{\hat N}_{\bf m} \;  
\eeq
is the total number operator of the bosons under investigation.

The Fock state $|n_{\bf m}\rangle$ represents the occupation number 
quantum state describing the number $n$ 
of bosons that are in the single-particle quantum state $|{\bf m}\rangle$. 
It satisfies the eigenvalue equation 
\beq 
{\hat N}_{\bf m} |n_{\bf m}\rangle = n_{\bf m} |n_{\bf m}\rangle \; . 
\eeq

We can also introduce a generic multi-mode Fock state \cite{salasnich2017quantum,crisanti2019condensation} 
of our problem associated with
the set of occupation numbers 
\beq
{\bf n} = \{n_{\bf m}\} = (n_{00},n_{01},n_{10},n_{11}, \ ...),
\label{P.0}
\eeq
as 
\beq 
|{\bf n}\rangle = \prod_{{\bf m}\in \mathbb{N}^2} |n_{\bf m}\rangle   
= |n_{00}\rangle \ |n_{01}\rangle \ |n_{10}\rangle \   |n_{11}\rangle  \ ...  
\eeq

This state is characterized by $n_{00}$ photons in the single-particle state $|00\rangle$, 
$n_{01}$ photons in the single-particle state $|01\rangle$, 
$n_{11}$ photons in the single-particle state $|11\rangle$, et cetera. 
These multi-mode Fock states can be used to obtain 
the following spectral resolution of the identity
\beq 
{\hat 1} = \sum_{\bf n} 
|{\bf n}\rangle \langle {\bf n} | \; .
\label{58}
\eeq

\subsection{Grand Canonical Formulation}

In order to describe the phenomenon of the Bose--Einstein condensation in this
system, we start from the GCE, where the density operator is
\beq
\hat{D}=e^{-\beta ({\hat H}-\mu {\hat N})},
\eeq
and the probability of the set of occupation numbers ${\bf n}$
is given by
\beq
P_\textrm{GC}({\bf n})  =  \frac{1}{{Z}_\textrm{GC}}  \langle {\bf n}|\hat{D}|{\bf n}\rangle= \frac{1}{{Z}_\textrm{GC}} e^{-\beta \sum_{\bf m} (\epsilon_{\bf m} -\mu )n_{\bf m}},
\label{P.1}
\eeq
where  
\beq 
{Z}_\textrm{GC}  =   
Tr[e^{-\beta ({\hat H}-\mu {\hat N})}]  = 
\prod_{\bf m} \left [ 1- e^{-(\beta \lambda_{\bf m} + \kappa)} \right ]^{-1} 
\eeq
and 
\beqa
\label{P.2} \\
& & \lambda_{\bf m}  =  \epsilon_{\bf m} - \epsilon_{\bf 0},
\label{P.3} \\
& & \kappa  =  \beta(\epsilon_{\bf 0} - \mu). 
\label{P.4}
\eeqa

\subsubsection{Equation of State}

The grand canonical 
thermal average of the total number operator reads
\beq
\langle {\hat N} \rangle_{GC} = \frac{Tr[{\hat N} e^{-\beta ({\hat H}-\mu {\hat N})}]}{{Z}_{GC}} 
= - \frac{\partial}{\partial \kappa} \ln {Z}_\textrm{GC},
\label{P.5}
\eeq
where
\beq
-\ln {Z}_\textrm{GC} = \sum_{\bf m} \, \ln \left [ 1- e^{-(\beta \lambda_{\bf m} + \kappa)} \right],
\label{P.7}
\eeq
and
\beq
\langle {\hat N}_{\bf m}  \rangle = \frac{1}{e^{(\beta \lambda_{\bf m} + \kappa)} - 1}.
\label{P.8}
\eeq

In order to proceed further, we now introduce the definition of the thermodynamic limit for our system of bosons confined by a harmonic potential. Following~\cite{dalfovo1999theory}, we consider the~conditions
\be
\label{eq:Th_Lim1}
  N\gg1, \quad \omega \ll 1, \qquad N \hbar^2\omega^2 = \rho = \text{finite},
  \ee
  where we have included $\hbar$ into the definition of the pseudo-density $\rho$ for simplicity. To keep the notation formally similar to the more familiar case of a system of particles in a box of volume $V$, as considered in the previous sections, we now introduce a pseudo-volume $V=1/(\hbar\omega)^2$ so that the condition of the thermodynamic limit can be also written as $N\gg1$ and $V\gg 1$, with $\rho=N/V$ fixed.
  
Next, separating from the sum in Equation~(\ref{P.7}) the ${\bf m} = 0$ term, we have
\be
\label{eq:ZGC_ex}
  \ln Z_\GCe(\beta,s) = -\ln\bigl(1 - s\bigr) - \frac{V}{\beta^2} J(s),
\ee
where we have defined $s = e^{-\kappa} = z\E^{-\beta \epsilon_{\bf 0}}\leq 1$, which plays the role of a fugacity in a rescaled reference frame with zero lowest energy. In the following, we will find it more convenient to consider the thermodynamic quantities as a function of $s$ rather than $z$.  
The function {\color{black} $J(s)\equiv -g_3(s)$, where $g_n(s)$ denotes the Bose functions~\cite{ziff1977ideal}}, represents the sum over the excited states and is given by
\be
\label{eq:Jz}
  J(s) = \int_{0}^{+\infty} dy\, y\, \ln\bigl(1- s\,\E^{-y}\bigr).
\ee

Using  \eqref{eq:ZGC_ex}, we have
\be
\label{eq:Nav}
\langle \hat{N}\rangle_\GCe =  s\frac{\partial}{\partial s} \ln Z_\GCe(\beta,s) 
                                    = \frac{s}{1-s} + N_1(\beta,s),
\ee
where
\be
\label{eq:Nb}
  N_1(\beta,s) = \frac{V}{\beta^2} I(s),
\ee
with
\be
\label{eq:Is}
  I(s) {\color{black}\equiv g_2(s)}= -s\frac{\partial}{\partial s} J(s) =  \int_{0}^{+\infty} dy\, \frac{y}{s^{-1}\,\E^{y}-1}.
\ee

\subsubsection{Bose-Einstein Phase Transition}

The standard argument leading to the phenomenon of condensation is as follows: since $I'(s) > 0$,
the function $I(s)$ is a monotonous increasing function of $s$,
so that, for any $\beta$, one has
\be
\label{eq:Nc}
  N_1(\beta,s) \leq N_{\rm c}(\beta) = \frac{V}{\beta^2} I(1).
\ee

If $I(1)$ is finite, as in our case  $I(1) = \pi^2/6$,  $N_{\rm c}(\beta)$ is a finite
 decreasing function of $\beta$ so that, for any fixed integer number $N$, there exists a finite $\beta_{\rm c}(N)$, the inverse critical temperature,
 such that
\be
\label{eq:bc_N}
   \beta_{\rm c}(N): \quad N_{\rm c}(\beta_{\rm c}) = N.
   \ee

   To address the issue of the ensemble {(in)equivalence} in the condensed phase, we consider 
   the condition
   \be
   \label{eq:condition}
   \langle \hat{N}\rangle_\GCe =N,
   \ee
   where $N$ now plays the role of a control parameter in the corresponding CE (see also the derivation reported in Appendix B leading to Equation~(\ref{eq:Nave_Nlarge})).
Then, for $N > N_{\rm c}(\beta)$, or $\beta > \beta_{\rm c}(N)$,
the contribution from the lowest energy level in 
Equation~\eqref{eq:Nav} cannot be neglected.
In terms of the pseudo-density $\rho$, the conditions \eqref{eq:Nc} and \eqref{eq:bc_N} become
\be
\label{eq:rho_bar}
     \rho_1(\beta,s) \leq \rho_{\rm c}(\beta) = \frac{1}{\beta^2} I(1),
\ee
and
\be
\label{eq:bc_rho}
   \beta_{\rm c}(\rho): \quad \rho_{\rm c}(\beta_{\rm c}) = \rho.
\ee

Then, from the condition~\eqref{eq:condition}, one has
\be
\label{eq:rho_z}
  \rho = \frac{1}{V}\frac{s}{1-s} + \rho_1(\beta,s) , 
  \ee
{where 
$\rho_1(\beta,s)=N_1(\beta,V,s)/V$. 
We stress that, in Equation (\ref{eq:bc_rho}), the density $\rho$ is an arbitrary parameter, as $N$ in Equation (\ref{eq:bc_N}). In Equation (\ref{eq:rho_z}), the density $\rho$ on the left-hand side refers to
the canonical ensemble $(\beta, V, N )$ while the quantities on the right-hand side refer to the
grand canonical ensemble $(\beta,V,s)$}. 

From this structure, it is clear that, in order to study the thermodynamic limit correctly,
we must distinguish two regions: the first, where 
$1-s = O(1)$ as $V\gg 1$, which is appropriate for $\beta < \beta_{\rm c}$, and the second, where $1-s = O(1/V)$ as 
$V\gg 1$, describing the case $\beta > \beta_{\rm c}$. {In singular perturbation theory, this case represents an instance of a {\sl boundary layer problem} with a  {\sl boundary layer} at $s=1$. The regions are called,  respectively, the {\sl outer} and {\sl inner} region in boundary layer theory.}
We note that the results discussed in Section \ref{sec:Th_Lim} hold for $\beta < \beta_{\rm c}$ above the critical temperature. 

\subsection{{\color{black}The Kac Kernel}}
\label{Kac}

We now study the behavior of the Kac kernel, {which represents the mathematical transformation connecting corresponding quantities in the two ensembles. This allows us to address the issue} related to the ensemble equivalence, in the case $\beta > \beta_{\rm c}$. {\color{black} The explicit form of the kernel was first derived in~\cite{ziff1977ideal} in the case of a uniform system. Here, we obtain the same results for the case of the trapped gas following an alternative approach. In particular,} we have to consider what happens 
in the region where $1-s = O(1/V)$ as 
$V\gg 1$, and this can be achieved by performing a convenient change in variable prior to the thermodynamic limit.
To be specific, let us rewrite $K(N|z)$ in the following form (see Equation~\eqref{eq:KNz-GC} in the Appendix A for details):
\be
\label{solo-lui-bis}
  K(N'|z)  = \frac{1}{2\pi i} \oint_{\Gamma} d\xi\, \xi^{-1-N'} \, 
     \frac{Z_\GCe(\beta,\xi z)}{Z_\GCe(\beta,z)}.
\ee

We use $N'$ to emphasize  that it is a running argument unrelated to $z$. {The fugacity $z$ is related to $\langle {\hat N}\rangle_{GC}$} 
imposing the condition \eqref{eq:rho_z} 
{but substituting $\rho$ with $\bar{\rho}=\langle {\hat N}\rangle_{GC}/V$, namely 
\beq 
\label{eq:rho_av_z}
\bar{\rho} = \frac{1}{V}\frac{s}{1-s} + \rho_1(\beta,s).
\eeq
} 

We introduce the variable $\eta$ through the relation  
$s = z\E^{-\beta\eps_{\bf 0}} = 1- \eta/V$, where $\eta = O(1)$ as $V\gg 1$. A similar change in variable must be carried out for $\xi$.
We have seen in Section \ref{sec:Th_Lim} that, in the limit of large $N$, the integral is dominated  
by a path parallel to the imaginary axis. This remains true also for $\beta > \beta_{\rm c}$ because
$(\partial/\partial\xi)^2 \ln Z_\GCe(\beta,\xi z) > 0$. Thus, we write 
$\xi = 1 + iv/V$, where $v = O(1)$ as $V\gg 1$.
Using \eqref{eq:ZGC_ex} and
\be
  \xi s = (1+iv/V)(1-\eta/V) \sim 1 - (\eta - iv)/V + O(1/V^2), 
\ee
we have
\be
\begin{split}
  \ln \frac{Z_\GCe(\beta,\xi z)}{Z_\GCe(\beta,z)} &\sim
     -\ln\bigl( (\eta - iv)/V\bigr)  + \ln\bigl(\eta/V\bigr) \\
    & - \frac{V}{\beta^2} \Bigl[ J(1-(\eta - iv)/V ) -  J(1-\eta/V ) \Bigr]
     \\
     & \sim
     - \ln \bigl(1 - iv/\eta) - J'(1) \frac{iv}{\beta^2} + O(1/V)
     \\
     & \sim
     - \ln \bigl(1 - iv/\eta) + i\rho_{\rm c}v + O(1/V),
\end{split}     
\ee
because, from Equations~\eqref{eq:Is} and \eqref{eq:rho_bar}, it follows that $J'(1)/\beta^2 = -\rho_{\rm c}$.
Using now
\be
  \xi^{-N'} = \E^{-N'\ln(1+iv/V)}\sim \E^{-i\rho'v},
\ee
where $\rho'=N'/V$, we have
\be
\label{eq:Kz_crt}
\begin{split}
  K(N'|z) &\sim \frac{1}{2\pi V} \int_{-\infty}^{\infty} dv\, \frac{\E^{-i(\rho'-\rho_{\rm c})v}}{1 - iv/\eta}
  \\
     &\sim -\frac{\eta}{2\pi i V} \int_{-\infty}^{\infty} dv\, \frac{\E^{-i(\rho'-\rho_{\rm c})v}}{v +  i\eta} \\
    &\sim \Delta\rho'\, \eta\, \E^{-(\rho'-\rho_{\rm c})\, \eta}\, \theta(\rho'-\rho_c),
\end{split}  
\ee
with $\Delta\rho' = 1/V$.
This expression can be written in terms of $\rho = N/V$, expressing 
$\eta = V ( 1 - z\E^{-\beta\eps_0})$ and  using Equation~\eqref{eq:rho_z}. Indeed,
substituting $s=1-\eta/V$ into Equation~\eqref{eq:rho_z}, we obtain
\be
\label{eq:eta_rho}
  \rho = \frac{1-\eta/V}{\eta} + 
  {{\rho}_1}(1-\eta/V,\beta)
         \sim \eta^{-1} + \rho_{\rm c}(\beta) + O(1/V),
\ee
and replacing $\eta^{-1} = \rho -\rho_{\rm c}$ in Equation~\eqref{eq:Kz_crt},
we finally have the explicit form of the kernel:
\be
\label{eq:Krr}
K(\rho'|\rho) \equiv K(N'|z)  / \Delta\rho'  = \frac{\E^{-(\rho'-\rho_{\rm c})/(\rho - \rho_{\rm c})}}{\rho -\rho_{\rm c}}\,
                    \theta(\rho'-\rho_{\rm c}),
\ee
valid for $\beta > \beta_{\rm c}$.
This result was obtained by Ziff {et al.} for the free ideal Bose gas in \cite{ziff1977ideal}, and the kernel $K(\rho'|\rho)$ is known as Kac density. 
{ In Appendix C, we give an alternative derivation of Equations (\ref{eq:Kz_crt}) and (\ref{eq:Krr}) that makes use of the Laplace method.} 

The explicit expression of the kernel can be used to connect averages in the CE and GCE expressed in terms of $\rho$
\be
\label{ziff0}
\langle f\rangle_{GC}(\rho)=\int_0^\infty d\rho' \, K(\rho'|\rho) \, \langle f\rangle_{C}(\rho').
\ee

In particular, it is easy to show that
$\int d\rho'\, K(\rho'|\rho)=1$, and that
the following relation holds:
\be
\label{ziff}
(\rho-\rho_c)^k=\int_0^\infty d\rho' \, K(\rho'|\rho)\frac{(\rho'-\rho_c)^k}{k!}.
\ee

The above result will play an important role in the study of the occupation number fluctuations
in the canonical and grand canonical ensembles.

Finally, we observe that, when expressed in the variable $z$, the kernel takes the form
\be
  K(N'|z) / \Delta\rho' = V(1-s)\, \E^{-V(1-s)(\rho' - \rho_{\rm c})},
\ee
and therefore, in the limit $V\gg 1$ with $1-s= O(1)$, i.e., for $\eta \gg 1$, it reduces to
\beq
  K(N'|z) = \Delta\rho' \delta(\rho'-\rho_{\rm c}) = {\delta^{Kr}(N'-N_{\rm c})} , 
\eeq
{where $\delta^{Kr}$ is the Kronecker delta.} 
Thus, as expected, in the limit $\beta\to \beta_{\rm c}^{-}$, we recover the expression
\eqref{eq:KNz_dir} valid for $\beta < \beta_{\rm c}$, i.e., in the non-condensed phase.

\subsection{Density Fluctuations and Grand Canonical Catastrophe}

As mentioned before, the explicit form of the Kac kernel allows us to connect averages in the GCE with those in the CE. {\color{black} Since the analytical expression of the Kac kernel is the same as in the case of the uniform system, the relations between density fluctuations in the CE and in the GCE coincide with those derived in~\cite{ziff1977ideal}. We report here the formulae to keep the paper self-contained.}
In particular, the average number of bosons in the condensed phase in the GCE is given by Equation (\ref{P.8}) and therefore
one has, for large $V$,
\begin{equation}
\langle \hat{N}_{\bf 0}\rangle_{GC} = \frac{1}{s^{-1}-1}\sim V(\rho-\rho_c),  
\label{ia.2}
\end{equation}
where we have used Equation \eqref{eq:eta_rho}, and the dependence on $T$ is through $\rho_c$. 
From the relations (\ref{ziff0}) and (\ref{ziff}), we then obtain that the averages of $\hat{N}_{\bf 0}$ are equal in the two ensembles
\begin{equation}
\langle \hat{N}_{\bf 0}\rangle_{C}=\langle \hat{N}_{\bf 0}\rangle_{GC}.
\end{equation}

On the contrary, considering the mean square occupation number, in the GCE, one has
\begin{equation}
\langle \hat{N}_{\bf 0}^2\rangle_{GC}=\langle \hat{N}_{\bf 0}\rangle_{GC}+2\langle \hat{N}_{\bf 0}\rangle_{GC}^2=V(\rho-\rho_c)+2V^2(\rho-\rho_c)^2
\end{equation}
and therefore
\begin{equation}
\langle \hat{N}_{\bf 0}^2\rangle_{GC}-\langle \hat{N}_{\bf 0}\rangle_{GC}^2=V(\rho-\rho_c)+V^2(\rho-\rho_c)^2.
\end{equation}

Exploiting again the Relation (\ref{ziff}),
in the CE, one obtains 
\begin{equation}
\langle \hat{N}_{\bf 0}^2\rangle_{C}= \langle {\hat N}_{\bf 0}\rangle_C + \langle{\hat N}_{\bf 0}\rangle_C^2=V(\rho-\rho_c)+V^2(\rho-\rho_c)^2,
\end{equation}
and 
\begin{equation}
\langle \hat{N}_{\bf 0}^2\rangle_{C}-\langle \hat{N}_{\bf 0}\rangle_{C}^2=V(\rho-\rho_c).
\end{equation}

In particular, for the second-order correlation function (at zero time delay; {see also Section \ref{sec341}}),
we have
\beq 
{ g^{(2)}(0) \equiv} 
\frac{\langle {\hat N}_0^2\rangle-\langle {\hat N}_0\rangle}{\langle{\hat N}_0\rangle^2}
= 
\left\{ 
\begin{matrix} 
1 & \mbox{in the CE} \\
2 
& \mbox{in the GCE,} \\
\end{matrix}
\right. 
\label{g2}
\eeq
in agreement with what was observed in experiments~\cite{klaers2012statistical}. {Indeed, in Ref. \cite{klaers2012statistical} (see {Figure 2b}), experimental results are reported showing that $g^{(2)}(0)=2$, even in the low-temperature phase, if a larger reservoir size is considered,  which corresponds to the realization of a genuine grand canonical conditions. On the other hand, when canonical conditions are realized, the same figure shows that fluctuations are damped, and $g^{(2)}(0)=1$ is observed. } 

Moreover, for the intensive density fluctuations, one has
\begin{equation}\label{CEfluct}
\lim_{V\to\infty}\frac{\langle \hat{N}_{\bf 0}^2\rangle_{C}-\langle \hat{N}_{\bf 0}\rangle_{C}^2}{V^2}=0 \;\; \textrm{in the CE},
\end{equation}
while
\begin{equation}\label{GCEfluct}
\lim_{V\to\infty}\frac{\langle \hat{N}_{\bf 0}^2\rangle_{GC}-\langle \hat{N}_{\bf 0}\rangle_{GC}^2}{V^2}=(\rho-\rho_c)^2 \;\; \textrm{in the GCE}.
\end{equation}

The latter result is known as a grand canonical catastrophe~\cite{ziff1977ideal} because of the counter-intuitive phenomenon of macroscopic fluctuations,
non-vanishing in the low temperature limit $T\to 0$.
This finding shows the inequivalence of ensembles at the level of fluctuations and
has been interpreted as an example where the GCE is not appropriate for computing averages~\cite{ziff1977ideal,holthaus1998condensate,fujiwara1970fluctuations,kocharovsky2006fluctuations,yukalov2007bose}.
However, in the experimental setup with photons in the microcavity, where GCE conditions are realized,  
macroscopic fluctuations have been actually observed~\cite{schmitt2018dynamics}, 
making the prediction of the GCE physically consistent~\cite{zannetti2015grand}.

The mathematical origin of such huge fluctuations can be traced back to the specific form of the Kac kernel
connecting the CE and GCE, {\color{black} as detailed in the previous subsection}. 

\subsubsection{Spatial Density--Density Correlation Function of the Condensate}\label{sec341}

We now discuss another quantity that is affected by the grand canonical catastrophe: 
the spatial density--density correlation function of the Bose--Einstein condensate. 

We start by noting that, for both the CE and GCE, 
the one-body correlation function 
$\langle {\hat \psi}^+({\bf r}) {\hat \psi}({\bf r}') \rangle $ 
can be written in terms of ladder operators in this way: 
\beqa 
\langle {\hat \psi}^+({\bf r}) {\hat \psi}({\bf r}') \rangle 
&=& \sum_{{\bf m},{\bf m}'} 
\langle {\hat a}^+_{\bf m} {\hat a}_{{\bf m}'} \rangle 
\phi^*_{\bf m}({\bf r}) \phi_{{\bf m}'}({\bf r}') 
\nonumber 
\\
&=& \langle {\hat a}^+_{\bf 0} {\hat a}_{{\bf 0}} \rangle 
\phi_{\bf 0}^*({\bf r}) \phi_{{\bf 0}}({\bf r}') + ... 
\nonumber 
\\
&=&  \langle {\hat N}_{\bf 0} \rangle 
\phi_{\bf 0}^*({\bf r}) \phi_{{\bf 0}}({\bf r}') + ... \; ,  
\eeqa
where the dots represent the contributions of the excited states. 

Working at zero temperature, where the effect due to the excited states 
goes to zero, we find that
\beq 
\langle {\hat \psi}^+({\bf r}) {\hat \psi}({\bf r}') \rangle 
= \langle {\hat N}_{\bf 0} \rangle 
\phi_{\bf 0}^*({\bf r}) \phi_{{\bf 0}}({\bf r}') \; . 
\eeq

Previously, we have seen that, at zero temperature,  
\beq 
\langle {\hat N}_{\bf 0}\rangle = N
\eeq
for both the CE and GCE.
Thus, 
\beq 
\langle {\hat \psi}^+({\bf r}) {\hat \psi}({\bf r}') \rangle = 
N \, \phi_{\bf 0}^*({\bf r}) \phi_{{\bf 0}}({\bf r}') \; . 
\eeq

Recalling Equation (\ref{condwave}), namely 
$\phi_{\bf 0}({\bf r})=e^{-|{\bf r}|^2/(2l_{H}^2)}/(\pi^{1/2}l_{H})$ 
with $l_{H}=\sqrt{\hbar/(m_{eff}\omega)}$, we eventually obtain 
\beq 
\langle {\hat \psi}^+({\bf r}) {\hat \psi}({\bf r}') \rangle  = 
{N\over \pi l_H^2} \, e^{-(|{\bf r}|^2+|{\bf r}'|^2)/(2l_H^2)} \; . 
\eeq

Setting ${\bf r}'={\bf 0}$, the previous expression becomes 
\beq 
\langle {\hat \psi}^+({\bf r}) {\hat \psi}({\bf 0}) \rangle = 
{N\over \pi l_H^2} \, e^{-|{\bf r}|^2/(2l_{H}^2)},    
\eeq
while, setting ${\bf r}'={\bf r}$, we obtain the result 
\beq 
\langle {\hat \psi}^+({\bf r}) {\hat \psi}({\bf r}) \rangle = 
\langle {\hat \rho}({\bf r}) \rangle = 
{N\over \pi l_H^2} \, e^{-|{\bf r}|^2/l_{H}^2} \; .   
\eeq

We then study the density--density
correlation function $\langle {\hat \rho}({\bf r}) {\hat \rho}({\bf
  r}') \rangle $, which, for both the CE and GCE, can be written in terms of
ladder operators as follows:
\beqa \langle {\hat \rho}({\bf r}) {\hat
  \rho}({\bf r}') \rangle &=& \sum_{{\bf m},{\bf m}',{\bf m}'',{\bf
    m}'''} \langle {\hat a}^+_{\bf m} {\hat a}_{{\bf m}'} {\hat
  a}^+_{{\bf m}''} {\hat a}_{{\bf m}'''} \rangle \nonumber \\ &\times&
\phi^*_{\bf m}({\bf r}) \phi_{{\bf m}'}({\bf r}) \phi^*_{{\bf
    m}''}({\bf r}') \phi_{{\bf m}'''}({\bf r}') \nonumber \\ &=&
\langle {\hat a}^+_{\bf 0} {\hat a}_{{\bf 0}} {\hat a}^+_{{\bf 0}}
        {\hat a}_{{\bf 0}} \rangle |\phi_{\bf 0}({\bf r})|^2
        |\phi_{{\bf 0}}({\bf r}')|^2 + ...  \nonumber \\ &=& \langle
        {\hat N}_{\bf 0}^2 \rangle |\phi_{\bf 0}({\bf r})|^2
        |\phi_{{\bf 0}}({\bf r}')|^2 + ... \; ,
        \eeqa
        where the dots
        stand for the contributions of the excited states.  Similarly,
        we have
        \beqa \langle {\hat \rho}({\bf r})\rangle \langle
        {\hat \rho}({\bf r}') \rangle &=& \sum_{{\bf m},{\bf m}',{\bf
            m}'',{\bf m}'''} \langle {\hat a}^+_{\bf m} {\hat a}_{{\bf
            m}'} \rangle \langle {\hat a}^+_{{\bf m}''} {\hat a}_{{\bf
            m}'''} \rangle \nonumber \\ &\times& \phi^*_{\bf m}({\bf
          r}) \phi_{{\bf m}'}({\bf r}) \phi^*_{{\bf m}''}({\bf r}')
        \phi_{{\bf m}'''}({\bf r}') \nonumber \\ &=& \langle {\hat
          a}^+_{\bf 0} {\hat a}_{{\bf 0}} \rangle \langle {\hat
          a}^+_{{\bf 0}} {\hat a}_{{\bf 0}} \rangle |\phi_{\bf 0}({\bf
          r})|^2 |\phi_{{\bf 0}}({\bf r}')|^2 + ...  \nonumber \\ &=&
        \langle {\hat N}_{\bf 0} \rangle^2 |\phi_{\bf 0}({\bf r})|^2
        |\phi_{{\bf 0}}({\bf r}')|^2 + ... \; .
        \eeqa
        
        Thus,
        considering the zero temperature limit for simplicity and neglecting
        the effect due to the excited states, we obtain
        \beqa \langle
        {\hat \rho}({\bf r}) {\hat \rho}({\bf r}') \rangle - \langle
        {\hat \rho}({\bf r})\rangle \langle {\hat \rho}({\bf r}')
        \rangle &=& \left( \langle {\hat N}_{\bf 0}^2 \rangle - \langle
                {\hat N}_{\bf 0} \rangle^2 \right) \nonumber \\
                &\times& |\phi_{\bf 0}({\bf
                  r})|^2 |\phi_{{\bf 0}}({\bf r}')|^2 \; .  \eeqa
              
By using our previous results of Equations (\ref{CEfluct}) and (\ref{GCEfluct}), 
which crucially depend on the statistical ensemble, 
at zero temperature, we then have 
\beq 
{\langle {\hat N}_{\bf 0}^2 \rangle - 
\langle {\hat N}_{\bf 0} \rangle^2 \over V^2} = 
\left\{ 
\begin{matrix} 
0 & \mbox{ in the CE} \\
\rho^2 & \mbox{ in the GCE.} \\
\end{matrix}
\right. 
\label{thvsexp1}
\eeq 

As a consequence, for the density--density correlation function \emph{per particle},
we finally obtain 
\beq 
\frac{\langle {\hat \rho}({\bf r}) {\hat \rho}({\bf r}') \rangle 
- \langle {\hat \rho}({\bf r})\rangle \langle {\hat \rho}({\bf r}') \rangle}{N} =0
\eeq
in the CE, 
and 
\beq 
\frac{\langle {\hat \rho}({\bf r}) {\hat \rho}({\bf r}') \rangle 
- \langle {\hat \rho}({\bf r})\rangle \langle {\hat \rho}({\bf r}') \rangle}{N} =
{N\over \pi^2 l_H^4}
e^{ -(|{\bf r}|^2+|{\bf r}'|^2)/l_{H}^2},
\label{thvsexp2}
\eeq
in the GCE. Equation (\ref{thvsexp2}) is not translationally invariant, 
as expected, due to the presence of the confining harmonic potential. 
Moreover, setting ${\bf r}'={\bf 0}$, we obtain 
\beq 
\frac{\langle {\hat \rho}({\bf r}) {\hat \rho}({\bf 0}) \rangle 
- \langle {\hat \rho}({\bf r})\rangle \langle {\hat \rho}({\bf 0}) \rangle}{N} 
= 
\left\{ 
\begin{matrix} 
0 & \mbox{ in the CE} \\ \\
{N\over \pi^2 l_H^4} 
e^{-|{\bf r}|^2/l_{H}^2}
& \mbox{ in the GCE}, \\
\end{matrix}
\right. 
\eeq
showing a decay to zero for a large $|{\bf r}|$ with a correlation  
length $l_H$, representing the size of the bosonic cloud, that diverges in the thermodynamic limit as $\omega\to 0$.
Note that the prefactor in the above expression remains finite in the thermodynamic limit.

\subsection{Internal Energy and Specific Heat}

To complete the description of the thermodynamic properties of the
system, we compute the internal energy and the specific heat in the normal and
condensed phases. These quantities have been measured in experiments
in~\cite{damm2016calorimetry}, where a comparison with a numerical
evaluation of the Bose--Einstein distribution function is
reported. Here, we provide close analytical expressions for average
energy and specific heat, valid both above and below the critical
temperature (see also ref.~\cite{grossmann1995lambda} for a similar
computation in a 3D~system).

\subsubsection{Grand Canonical Ensemble}

The (average) internal energy of the system in the GCE is given by
\be
\begin{split}
  E_{\GCe}(\beta,z) &= -\pder{\beta}{z}\ln Z_\GCe(\beta,z)
              \\
                   &= \pder{s}{\beta}\left[\ln\bigl(1 - s\bigr) + \frac{1}{(\beta \hbar \omega)^2} J(s)\right]\,
                     \pder{\beta}{z}s    \\              
                     &+ \pder{\beta}{s} \frac{1}{(\beta \hbar \omega)^2} J(s)
                     \\
                     &=\ov{N}(\beta,z)\,\eps_0 - 2\frac{\beta^{-3}}{(\hbar\omega)^2}\, J(s),
\end{split}  
\ee
where we have used
\be
  \pder{\beta}{z}s = -\eps_0\, s,
\ee
and
\beqa
   &-&s\,\pder{s}{\beta}\left[\ln\bigl(1 - s\bigr) + \frac{1}{(\beta \hbar \omega)^2} J(s)\right]
   =\nonumber \\
   &=&z\,\pder{z}{\beta} \ln Z_\GCe(\beta,z) = \ov{N}(\beta,z) = \langle \hat{N}\rangle_{\GCe}.
\eeqa

{{Note that zero-point energy goes to zero in the thermodynamic limit. Here, we keep this term because we take into account first-order corrections.}}

Evaluating the derivative, we have the explicit form
\be
\label{eq:N_av}
   \ov{N}(\beta,z) = \frac{s}{1-s} + \frac{1}{(\beta\hbar\omega)^2}I(s).
\ee

{\color{black} In the above expressions, $J(s)$ and $I(s)$ are defined in Equations~(\ref{eq:Jz}) and (\ref{eq:Is}), respectively}.
Now using the condition $\langle \hat{N}\rangle_{\GCe} = N$, we have
\be
   E_{\GCe}(\beta,z) = N\,\eps_0  - 2\frac{\beta^{-3}}{(\hbar\omega)^2}\, J(s).
\ee

When $\beta > \beta_{\rm c}$, i.e., $T < T_{\rm c}$, the variable $s$ saturates to $s=1$, so we have
\be
\label{eq:E_bN}
  E_{\GCe}(\beta,N) = \begin{cases}
                     N\hbar\omega + 2\dfrac{\zeta(3)}{(\hbar\omega)^2}\, \beta^{-3}, & \beta > \beta_{\rm c}\\
                     \phantom{xx} \\
                     N\hbar\omega - 2\dfrac{J(s)}{(\hbar\omega)^2}\, \beta^{-3}, & \beta < \beta_{\rm c}\\                     
                     \end{cases}
\ee 
where we have used $J(1) = - \zeta(3)$.  In the regime $\beta < \beta_{\rm c}$, the variable 
$s\equiv s(\beta,N)$ is obtained from
\be
\label{eq:s_bN}
    \frac{1}{(\beta\hbar\omega)^2}I(s) = N,
    \qquad\Rightarrow\qquad I(s) = N(\beta\hbar\omega)^2.
\ee

Using
\be
  \frac{1}{\hbar\omega} = \sqrt{\frac{N}{I(1)}}\, \beta_{\rm c}
                                     = \sqrt{\frac{6N}{\pi^2}}\, \beta_{\rm c},
\label{old-mon}
\ee
we can express \eqref{eq:E_bN} in terms of $N$ and $T_{\rm c}$. A simple calculation leads to
\be
\label{eq:E_TN_1b}
  \frac{E_\GCe}{NT_{\rm c}} = \begin{cases}
                    \dfrac{\pi}{\sqrt{6}}\Bigr[ \dfrac{1}{N^{1/2}} + \dfrac{12\, \sqrt{6}}{\pi^3}\, \zeta(3)\, 
                    \bigl(T/T_{\rm c}\bigr)^{3}\Bigr], & T  < T_{\rm c}\\
                     \phantom{xx} \\
                     \dfrac{\pi}{\sqrt{6}} \Bigr[ \dfrac{1}{N^{1/2}} -  \dfrac{12\, \sqrt{6}}{\pi^3}\,J(s)\, 
                     \bigl(T/T_{\rm c}\bigr)^{3}\Bigr],
                     & T > T_{\rm c}\\                  
                     \end{cases}                    
\ee 
where $s=s(T,T_{\rm c})$ for $ T > T_{\rm c}$ is the solution of $I(s)=(\pi^2/6)(T_c/T)^2$.

Having the expression of the (average) internal energy, we can compute the specific heat.
From \eqref{eq:E_bN}, we have, for $\beta > \beta_{\rm c}$,
\be
  \pder{\beta}{N}E_{\GCe}(\beta,N) = -6\dfrac{\zeta(3)}{(\hbar\omega)^2}\, \beta^{-4}
\ee
which, using $(\partial/\partial T) = -\beta^2 (\partial/\partial\beta)$, leads to
\be
  \pder{T}{N}E_{\GCe}(T,N) = 6\dfrac{\zeta(3)}{(\hbar\omega)^2}
  \, T^{2}
                                      = 6N \dfrac{\zeta(3)}{I(1)}\, \left( \dfrac{T}{T_{\rm _c}} \right)^2,
\ee

Then,
\be
  c(T) = \frac{1}{N}\pder{T}{N}E_{\GCe}(T,N) = 6 \dfrac{\zeta(3)}{I(1)}\, \left( \dfrac{T}{T_{\rm _c}} \right)^2
\ee
or
\be
   c(T) = \dfrac{36\, \zeta(3)}{\pi^2}\, \left( \dfrac{T}{T_{\rm _c}} \right)^2,
   \qquad T < T_{\rm c}.
   \label{cgc-daversa1}
\ee

{This analytical result is in agreement with the very recent finding of ref. \cite{Paredes2024}.} 

For temperature $T {>} T_{\rm c}$, the calculation is a bit more involved. Again from 
\eqref{eq:E_bN}, we have, for $\beta < \beta_{\rm c}$,
\be
\begin{split}
  \pder{\beta}{N}E_{\GCe}(\beta,N) &= 6\dfrac{J(s)}{(\hbar\omega)^2}\, \beta^{-4}
                                             -2\dfrac{J'(s)}{(\hbar\omega)^2}\, \beta^{-3}\, \pder{\beta}{N}s
                                           \\
                                 &= 6\dfrac{J(s)}{(\hbar\omega)^2}\, \beta^{-4}
                                             +2\dfrac{I(s)}{s (\hbar\omega)^2}\, \beta^{-3}\, \pder{\beta}{N}s    
\end{split}                                                   
\ee
or, using \eqref{eq:s_bN}, 
\be
  \pder{\beta}{N}E_{\GCe}(\beta,N) = 6\dfrac{J(s)}{(\hbar\omega)^2}\, \beta^{-4}
                                             +2\dfrac{N}{s}\, \beta^{-1}\, \pder{\beta}{N}s.
\ee

Now from \eqref{eq:s_bN},  for $N=const$, we have
\be
  -2\dfrac{I(s)}{(\hbar\omega)^2}\beta^{-3}\, d\beta + \dfrac{I'(s)}{(\beta\hbar\omega)^2}\,  ds = 0,
\ee
and then
\be
  \pder{\beta}{N}s= 2\dfrac{I(s)}{I'(s)}\beta^{-1},
\ee
with
\be
\begin{split}
  I'(s) = \frac{d}{ds}\int_{0^+}^{+\infty}dy\, \frac{y}{s^{-1}\E^{y}-1}
         = \int_{0^+}^{+\infty}dy\, \frac{y\,\E^{-y}}{(1-s\,\E^{-y})^2}\\
         = - \frac{1}{s}\ln(1-s).
\end{split}         
\ee

Then,
\be
  \pder{\beta}{N}E_{\GCe}(\beta,N) = 6\dfrac{J(s)}{(\hbar\omega)^2}\, \beta^{-4}
                                             -4N \dfrac{I(s)}{\ln(1-s)}\, \beta^{-2},
\ee
which, with \eqref{eq:s_bN}, gives
\be
  \pder{T}{N}E_{\GCe}(T,N) = -6\dfrac{J(s)}{(\hbar\omega)^2}\, \beta^{-2}
                                             +4N^2 \dfrac{(\beta\hbar\omega)^2}{\ln(1-s)},
                                             \qquad T > T_{\rm c}.
\ee
so that, for $T>T_{\rm c}$,
\be
  c(T) = \frac{1}{N}\pder{T}{N}E_{\GCe}(T,N) =  -6\dfrac{J(s)}{N(\beta\hbar\omega)^2}
                                             +4N \dfrac{(\beta\hbar\omega)^2}{\ln(1-s)}.
\ee

If we finally express $\hbar\omega$ in terms of $N$ and $T_{\rm c}$, 
{i.e., using Equation (\ref{old-mon})}, we obtain
\be
\label{giafattotutto}
  c(T) = -6\frac{J(s)}{I(1)} \left(\frac{T}{T_{\rm c}}\right)^2 
            + 4\frac{I(1)}{\ln(1-s)}\left(\frac{T_{\rm c}}{T}\right)^2,
            \qquad T > T_{\rm c},
\ee
where $s=s(T,N)$ is again the solution of \eqref{eq:s_bN}. 
{Note that our Equation (\ref{giafattotutto}) also appears \linebreak in ref. \cite{Paredes2024}.}

\begin{figure}
    \includegraphics[width=8cm]{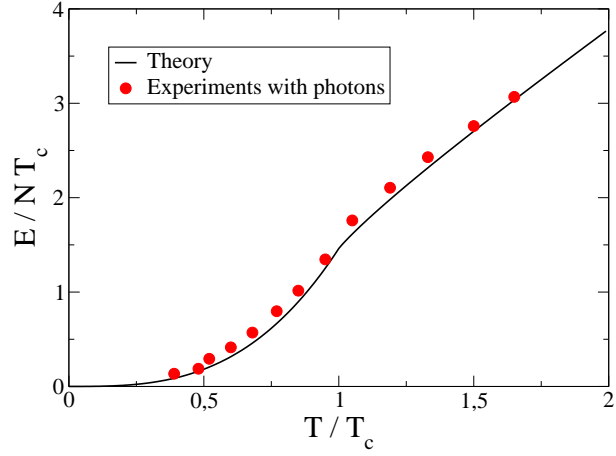}
    \caption{Internal 
 energy $E$ as a function of temperature $T$. Here, $T_c$ is the critical temperature of Bose--Einstein condensation. Solid line: our {analytical} theory in the grand canonical ensemble{, Equation (\ref{eq:E_TN_1b}) in the limit $N\to \infty$.} Filled circles: experimental data of ref. \cite{damm2016calorimetry}.}
    \label{fig:energy}
\end{figure}

\begin{figure}
    \includegraphics[width=8cm]{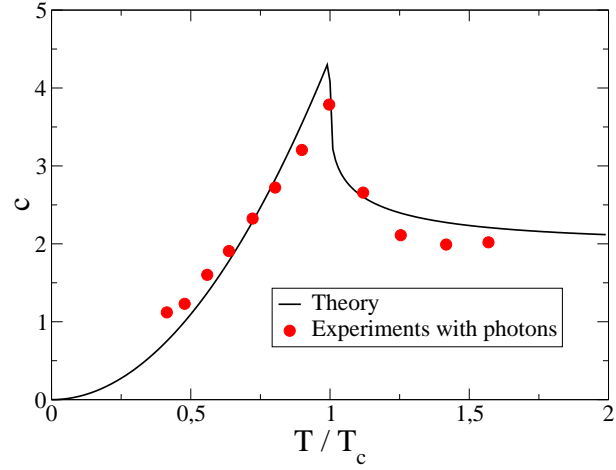}
    \caption{Specific  heat $c$ as a function of temperature. 
    $T_c$ is the critical temperature. Solid line: our  analytical results,
    {Equations (\ref{cgc-daversa1}) and (\ref{giafattotutto}) with $N\to\infty$,} in the grand canonical ensemble. Filled circles: experimental data of ref. \cite{damm2016calorimetry}.}
    \label{fig:cs}
\end{figure}

In Figures~\ref{fig:energy} and~\ref{fig:cs}, we plot the rescaled average energy $E/(N T_c)$ and the specific heat $c$ as a function of $T/T_c$, respectively, and compare the analytical predictions with experimental data taken from ref.~\cite{damm2016calorimetry}. 
There is a remarkable good agreement between our analytical results and the experimental ones. Notice that in Ref. \cite{damm2016calorimetry} are also reported theoretical curves, 
obtained with a numerical procedure, 
that are very similar to the solid curves of our Figures \ref{fig:energy} and \ref{fig:cs}. {The two figures strongly suggest that these experiments with photons are in a regime where the thermodynamic limit is practically achieved.} 

\subsubsection{Canonical Ensemble}

The same expression of the average internal energy, and therefore of the specific heat, also hold in the CE. Let us prove it. 

{\color{black} From Equation~\eqref{eq:ave}}, we know that 
\beqa
  \langle f\rangle_{\GCe}(s) &=& \sum_{N'\geq 0} \langle f\rangle_{\Ce}(N')\, K(N'|s)\nonumber \\
                      & = & \sum_{N'\geq 0} \Delta\rho'\,\langle f\rangle_{\Ce}(N')\, \dfrac{K(N'|s)}{\Delta\rho'},
\eeqa
where $\Delta\rho' = 1/V$; thus, in the thermodynamic limit, we have
\be
\label{eq:Ce_GCe}
  \langle f\rangle_{\GCe}(\rho) = \int_{0}^{+\infty} d\rho'\, \langle f\rangle_{\Ce}(\rho')\, K(\rho'|\rho),
\ee
where $s=s(\rho) = 1 - \eta/V$ with $\eta^{-1} = \rho -\rho_{\rm c}$,  and
\be
K(\rho'|\rho) \equiv K(N'|s)  / \Delta\rho'  
               = \frac{\E^{-(\rho'-\rho_{\rm c})/(\rho - \rho_{\rm c})}}{\rho -\rho_{\rm c}}\,
                    \theta(\rho'-\rho_{\rm c}).
\ee

From Equation~\eqref{eq:E_TN_1b}, neglecting the first term irrelevant for $N\to\infty$, we have
\be
   \dfrac{E}{N T_{\rm c}} = -\dfrac{2}{I(1)}\left(\dfrac{T}{T_{\rm c}}\right)^3\, J\bigl(s(\rho)\bigr).
\ee

Then, substituting
\be
\begin{split}
  J(1-\eta/V)&\sim J(1) - J'(1)\, \eta/V + O(1/V^2)
          \\
                  &\sim J(1) + I(1)\, \eta/V. + O(1/V^2),
                  \quad V\to \infty,
\end{split}                  
\ee
it follows that, in the {grand}  canonical ensemble,
\be
   \dfrac{E_\GCe}{N T_{\rm c}} \sim \frac{12}{\pi^2}\zeta(3) \left(\dfrac{T}{T_{\rm c}}\right)^3 
                                        -2 \left(\dfrac{T}{T_{\rm c}}\right)^3\, \frac{1}{V(\rho - \rho_{\rm c})} 
                                        + O(1/V^2),
\ee
where we have used $J(1) = -\zeta(3)$ and $I(1) = \pi^2/6$.

Using this expression in Equation \eqref{eq:Ce_GCe} with the explicit form of the kernel, {we obtain}
\be
\label{eq:egce}
  \begin{split}
         \int_{\rho_{\rm c}}^{+\infty} d\rho' \frac{\E^{-(\rho'-\rho_{\rm c})/(\rho - \rho_{\rm c})}}{\rho -\rho_{\rm c}}
             \dfrac{E_\Ce(\rho')}{N T_{\rm c}} \\
             \sim
     \dfrac{E_0}{N T_{\rm c}} -2 \left(\dfrac{T}{T_{\rm c}}\right)^3\, \frac{1}{V(\rho - \rho_{\rm c})} 
                                        + O(1/V^2)
  \end{split}
\ee
where
\be
    \dfrac{E_0}{N T_{\rm c}} = \frac{12}{\pi^2}\zeta(3) \left(\dfrac{T}{T_{\rm c}}\right)^3.
\ee

Now, using the normalization of $K(\rho'|\rho)$,
\be
  \int_{\rho_{\rm c}}^{+\infty} d\rho' \frac{\E^{-(\rho'-\rho_{\rm c})/(\rho - \rho_{\rm c})}}{\rho -\rho_{\rm c}} = 1,
\ee
it follows that
\be
  \dfrac{E_\Ce(\rho')}{N T_{\rm c}} \sim \dfrac{E_0}{N T_{\rm c}} 
                                          -2 \left(\dfrac{T}{T_{\rm c}}\right)^3\, \frac{1}{V}\,
                                           \delta(\rho'-\rho_{c}) +  O(1/V^2),          
\ee
{\color{black} as one can check by substituting back the above expression into Equation \eqref{eq:egce}.}
Recalling that $1/V=\Delta\rho'$ and using the identity
\be
\Delta\rho' \delta(\rho'-\rho_{\rm c}) = \delta^{\rm Kr}(N' - N_{\rm c}),
\ee
we have
\be
  \dfrac{E_\Ce(N')}{N T_{\rm c}} \sim \dfrac{E_0}{N T_{\rm c}} 
                                          -2 \left(\dfrac{T}{T_{\rm c}}\right)^3\, 
                                               V\,
                                           \delta^{\rm Kr}(N' - N_{\rm c}) +  O(1/V^2).
\ee

The number $N$ in the {grand} canonical ensemble and $N'$ in the canonical ensemble 
are  related by the requirement $N \equiv \langle \hat{N}\rangle_{\GCe} = N'$;
hence, the Kronecker delta vanishes and we conclude that
\be
  \dfrac{E_\Ce(N,T)}{N T_{\rm c}} =  \dfrac{E_\GCe(N,T)}{N T_{\rm c}} 
  =  \frac{12}{\pi^2}\zeta(3) \left(\dfrac{T}{T_{\rm c}}\right)^3,
  \qquad  T < T_{\rm c}.
\ee
in the thermodynamic limit. 

\section{Conclusions}
\label{Sec:Conc}

We have discussed a general formalism to derive physical quantities in the canonical ensemble 
from the corresponding ones in the grand canonical ensemble, where the calculations are usually much simpler. 
Then, {\color{black} motivated by recent experiments with photons}, we have applied this formalism to the study of an ideal Bose gas of particles under harmonic confinement in two 
spatial dimensions. Quite remarkably, also working in the thermodynamic limit, density fluctuations 
and spatial density--density correlations of the Bose--Einstein condensate display a strongly different 
behavior in the two ensembles. Similar to previous predictions for the uniform Bose gas \cite{ziff1977ideal},  
for the non-uniform condensate, we find that the density--density {correlation} is zero in the canonical 
ensemble and non-zero in the grand canonical ensemble. This result, which is known as the grand canonical 
catastrophe because of the counter-intuitive phenomenon of non-vanishing macroscopic fluctuations 
in the low temperature limit $T\to 0$, {turns out to be a real phenomenon as it has in fact been observed in experiments with photons in the microcavity~\cite{schmitt2018dynamics}. Our study sheds new light on the underlying mathematical and physical mechanisms that induce this intriguing behavior.}  
In the last part of this paper, we have also obtained analytical formulas for the internal energy and the specific heat 
both in the condensed phase and in the normal phase. 
{Similar results, obtained with a fully numerical procedure, can be found in ref. \cite{damm2016calorimetry}}.
{For these quantities, we have provided explicit analytical expressions, 
including also finite-size effects. The comparison with the experimental data} of ref. \cite{damm2016calorimetry}
shows a good agreement between our analytical theory and the empirical results. 

\vspace{6pt} 

L.S. is partially supported by the European Union-NextGenerationEU within 
the National Center for HPC, Big Data and Quantum Computing 
[Project No. CN00000013, CN1 Spoke 10: Quantum Computing], 
by the BIRD Project Ultracold atoms in curved geometries of the 
University of Padova, by Iniziativa Specifica Quantum of Istituto Nazionale 
di Fisica Nucleare, by the European Quantum Flagship Project PASQuanS 2,
by the PRIN 2022 Project Quantum Atomic Mixtures: Droplets, 
Topological Structures, and Vortices, and by the Project Frontiere Quantistiche within the 2023 funding programme  'Dipartimenti di Eccellenza' of the
Italian Ministry for Universities and~Research.
L.S. thanks Giacomo Gradenigo for useful discussions.

\section*{Appendix A}
                    
  By substituting the
  expression \eqref{eq:KNz}, we obtain \be \nonumber \langle
  f\rangle_\GCe(z) = \sum_{N} \langle f\rangle_\Ce(N)\, \frac{z^N\,
    Z_\Ce(N)}{Z_\GCe(z)}, \ee and hence \be \nonumber \langle
  f\rangle_\GCe(z)\, Z_\GCe(z) = \sum_{N} z^N\, Z_\Ce(N)\, \langle
  f\rangle_\Ce(N).  \ee 
  
  Inverting this relation gives \be \nonumber
  \langle f\rangle_\Ce(N)\, Z_\Ce(N) = \frac{1}{2\pi i}\,
  \oint_{\Gamma} dz\, z^{-1-N} \langle f\rangle_\GCe(z)\, Z_\GCe(z)
  \ee and hence \be
\label{eq:avC-avGC}
  \langle f\rangle_\Ce(N) = \frac{1}{2\pi i}\, \oint_{\Gamma} \frac{dz}{z} \langle f\rangle_\GCe(z)\, 
                      \frac{z^{-N} Z_\GCe(z)}{Z_\Ce(N)}.
\ee

Recalling the expression of $K(N|z)$, we finally obtain the following relation
\be
\langle f({\sigma})\rangle_\Ce(N) =  \frac{1}{2\pi i}\, \oint_{\Gamma} \frac{dz}{z} 
                    \frac{\langle f({\sigma})\rangle_\GCe(z)}{K(N|z)}
\ee
between averages in the GCE and averages in the CE. 

From the expression \eqref{eq:KNz}, which involves both $Z_\Ce(N)$ and $Z_\GCe(z)$, it seems that, in order to evaluate $K(N|z)$,
the partition functions in both ensembles should be explicitly known. However, {an alternative approach would be to} 
introduce the generating function 
\beqa
\label{eq:Kxiz}
\hat{K}(\xi|z) &=& \sum_{N} \xi^N\, K(N|z) 
                     = \frac{1}{Z_\GCe(z)} \sum_{N} \xi^N\, z^N\, Z_\Ce(N) \nonumber \\ 
                     &=& \frac{Z_\GCe(\xi z)}{Z_\GCe(z)},
\eeqa
so that
\beqa
K(N|z) &=& \frac{1}{2\pi i} \oint_{\Gamma} d\xi\, \xi^{-1-N} \hat{K}(\xi|z) \nonumber \\
           &=&  \frac{1}{2\pi i} \oint_{\Gamma} d\xi\, \xi^{-1-N} \frac{Z_\GCe(\xi z)}{Z_\GCe(z)}
\eeqa
is expressed in terms of $Z_\GCe(z)$ only.
This expression can also be obtained by substituting the
expression \eqref{eq:ZCe_ZGCe} into the definition \eqref{eq:KNz}:
\be
K(N|z) = \frac{z^N}{Z_\GCe(z)} \frac{1}{2\pi i} \oint_{\Gamma} ds\, s^{-1-N} \, Z_\GCe(s),
\ee
and setting $s=\xi z$
\be
\label{eq:KNz-GC}
K(N|z)  = \frac{1}{Z_\GCe(z)} \frac{1}{2\pi i} \oint_{\Gamma} d\xi\, \xi^{-1-N} \, Z_\GCe(\xi z).
 \ee

Finally, let us mention that the same result can also be obtained by introducing the discrete Fourier transform
\be
\label{eq:KFou}
   \overline{K}(\xi|z) = \sum_{N} \E^{iN\xi/V}\, K(N|z) 
                        = \frac{Z_\GCe(z\E^{i\xi/V})}{Z_\GCe(z)},
\ee
or, in terms of  the chemical potential $\mu$,
\be
   \overline{K}(\xi|\mu) = \frac{Z_\GCe(\mu + iT\xi/V)}{Z_\GCe(z)}.
\ee

In this case, it should be noted that 
\be
\label{eq:KFou1}
 K(N|z) = \int_{-\infty}^{+\infty} \frac{d\xi}{2\pi}\, \E^{-i\xi x}\,
 \overline{K}(\xi|z),
 \ee
 with $x=N/V$, is not the inverse of
 \eqref{eq:KFou}
 because
 \be \int_{-\infty}^{+\infty}
 \frac{d\xi}{2\pi}\, \E^{-i\xi x} = \delta(x),
 \ee
where $\delta(x)$ is the Dirac delta function.  If, however, the sum
 in \eqref{eq:KFou} is dominated by $N\gg1$, as in the thermodynamic
 limit, then $x=N/V$ can be considered a ``continuous'' variable and
 the sum can be replaced by an integral so that \eqref{eq:KFou1}
 becomes meaningful.

\section*{Appendix B}

Consider
\be
 Z_\Ce(N) = \frac{1}{2\pi i} \oint_{\Gamma}  dz z^{-1-N}\, Z_\GCe(z)
                 = \frac{1}{2\pi i} \oint_{\Gamma} \frac{dz}{z}\, \E^{N\psi(z)},
\ee
with
\be
\label{eq:phiz}
  \psi(z) = -\ln z + \frac{1}{N}\ln Z_\GCe(z).
\ee

In the limit $N\gg 1$, the integral is dominated by the {neighborhood} of the point 
$z^*$:
\be
  \psi'(z^*) = -\frac{1}{z^*} + \frac{1}{N} \frac{\partial}{\partial z^*} \ln Z_\GCe(z^*) = 0,
\ee 
which leads to the saddle point equation:
\be
\label{eq:GCe-Ce-Leg0}
z^* \frac{\partial}{\partial z^*} \ln Z_\GCe(z^*) = N.
\ee

Since $Z_\Ce(N)$ is real, the saddle point lies on the real axis, $\Im z^*= 0$.
A simple calculation then shows that
\beqa
  \psi''(z^*) &=& \frac{1}{{z^*}^2} + \frac{1}{N }\, \frac{\partial^2}{\partial z^2}\, \ln Z_\GCe(z) \Bigr|_{z=z^*}
        \nonumber \\
         &=& \frac{1}{{z^*}^2N }\, z^* \frac{\partial}{\partial z^*} \left[z^* \frac{\partial}{\partial z^*} \ln Z_\GCe(z^*)\right]
         \nonumber \\
         & = &\frac{1}{{z^*}^2N }\, \bigl[ \langle N^2\rangle_\GCe(z^*) - \langle N\rangle_\GCe(z^*)^2 \bigr]
         > 0.
\eeqa

Then, by expanding $\psi(z)$ about $z^*$ up to second order, as in Equation \eqref{eq:phir_2}, and
integrating along the steepest descent path parallel to the imaginary axis
$z=z^* + i y$, $ -\infty < y < +\infty$, we obtain
\be
\label{eq:ZCz}
  Z_\Ce(N) \sim \sqrt{\frac{\sigma_\psi^2}{2\pi}} \frac{\E^{N\psi(z^*)}}{z^*}
                 = \sqrt{\frac{\sigma_\psi^2}{2\pi}}\, {z^*}^{-1-N}\, Z_\GCe(z^*),
                 N\gg1,
\ee
where $z^*=z(N)$ is the solution of \eqref{eq:GCe-Ce-Leg0} and we have
defined $\sigma_\psi^{-2} = N\psi''(z^*) > 0$.
Dropping sub-leading terms as $N\gg1$, the expression \eqref{eq:ZCz} implies that
\be
\label{eq:GCe-Ce-Leg1}
  \ln Z_\Ce(N) = \ln Z_\GCe(z) - N \ln z, \qquad N\gg 1.
\ee
with $z=z^*= z(N)$ the solution of \eqref{eq:GCe-Ce-Leg0}.

{Equations \eqref{eq:GCe-Ce-Leg0} and \eqref{eq:GCe-Ce-Leg1} imply that,
  in the thermodynamic limit $N\gg 1$,
$\ln Z_\Ce(N)$ is the Legendre transform 
of $\ln Z_\GCe(z)$ with respect to $\ln z$, and hence
it is the inverse Legendre transform of  \eqref{eq:Ce-GCe-Leg0} and \eqref{eq:Ce-GCe-Leg1}.
Note also that \eqref{eq:GCe-Ce-Leg0} implies that, in the thermodynamic~limit,} 
\be
 \label{eq:Nave_Nlarge}
   \langle \hat{N}\rangle_\GCe(z^*) =  N = \langle \hat{N}\rangle_\Ce.
\ee

Finally, for the kernel $K(N|z)^{-1}$, we have
\be
    K(N|z)^{-1} = \frac{z^{-N}\, Z_\GCe(z)}{Z_\Ce(N)}  
                      = \E^{N\psi(z)}\left[\frac{1}{2\pi i} \oint_{\Gamma} \frac{dz}{z} \E^{N\psi(z)}\right]^{-1}.
\ee

In the thermodynamic limit, using \eqref{eq:ZCz}, we have
\be
  K(N|z)^{-1} \sim \sqrt{\frac{2\pi}{\sigma_\psi^2}}\, z^*\, \E^{\frac{1}{2\sigma^2_\psi}(z-z^*)^2}
                    = \sqrt{\frac{2\pi}{\sigma_\psi^2}}\, z^*\, \E^{-\frac{1}{2\sigma_\psi^2}\,y^2},
\ee
with $y=-i(z-z^*)$.  This expression can be simplified further by recalling that
$\sigma_\psi^{2} = O(1/N)$. Thus, 
\be
  K(N|z)^{-1} \sim  \frac{2\pi z^*}{\sqrt{2\pi \sigma_\psi^2}}\, \E^{-\frac{1}{2\sigma_\psi^2}\,y^2}
                    \sim 2\pi z^*\, \delta(y),
  \qquad N\gg 1,
\ee
and, using $\delta(ax) = (1/a)\delta(x)$, we finally have
\be
\label{eq:KNz_inv}
  K(N|z)^{-1}   \sim 2\pi i\, z\, \delta\bigl(z-z(N)\bigr),
  \qquad N\gg 1.
\ee
with $z(N)$ the solution of \eqref{eq:GCe-Ce-Leg1}.
Note that 
\be
   \frac{1}{2\pi i} \oint_{\Gamma} \frac{dz}{z}\, K(N,z)^{-1} = 1,
\ee
as follows from the definition of $K(N|z)$.

\section*{Appendix C}

In the derivation of the Kac kernel, the assumption $s = 1 -\eta/V$
may appear {ad hoc}.
Therefore, here, we present an alternative derivation of \eqref{eq:Kz_crt}. The starting point is Equation~(\ref{solo-lui-bis}), i.e., 
\be
\label{eq:KNz-GC_1}
  K(N|z) = \frac{z^{N}}{Z_\GCe(\beta,V,z)} \frac{1}{2\pi i} \oint_{\Gamma} d\xi\, \xi^{-1-N} \, 
             Z_\GCe(\beta, V, \xi) , 
\ee
where, here, we use $N$ instead of $N'$. The integral in this equation is  the canonical partition function (see \eqref{eq:ZCe_ZGCe}),
\be
  Z_\Ce(\beta,V, N) = \frac{1}{2\pi i} \oint_{\Gamma} d\xi \, \xi^{-1-N}\,  Z_\GCe(\beta,V, \xi),
\ee
where $\Gamma$ is a closed curve in the complex $\xi$-plane encircling the origin and 
not crossing the real axis on the cut $\Re\, \xi > \E^{\beta\eps_0}$.

If $\eps_0\not=0$, it is useful to define $s=\E^{-\beta\eps_0}z = \alpha_0 z$ so that
\be
\label{eq:ZC_a}
  Z_\Ce(\beta,V, N) = \frac{\alpha_0^N}{2\pi i} \oint_{\Gamma} ds \, s^{-1-N}\,  
         Z_\GCe(\beta, V, \alpha_0^{-1}s),
\ee
where, now, $\Gamma$ is a closed curve in the complex $s$-plane encircling the origin and 
not crossing the real axis on the cut $\Re\, s > 1$.

In our specific case, we have (see \eqref{eq:ZGC_ex})
\be
  Z_\GCe(\beta,V, \alpha_0^{-1}s) = -\frac{1}{1 - s}\,\E^{- \frac{V}{\beta^2} J(s)},
\ee
where $J(s)$ is given in \eqref{eq:Jz}.
Inserting this expression into \eqref{eq:ZC_a}, we have
\be
\label{eq:ZC_aa}
  Z_\Ce(\beta,V, N) = \frac{\alpha_0^N}{2\pi i} \oint_{\Gamma} 
    \frac{ds}{s} \, \frac{\E^{V\phi(s)}}{1-s},
\ee
where
\be
\label{eq:phi_a}
  \phi(s) = - \rho\,\ln s - \frac{1}{\beta^2}J(s),
\ee
where $\rho=N/V$ is the density in the canonical ensemble.

In the thermodynamic limit $V\to\infty$, $N\to\infty$ with $\rho = {\text constant}$, the integral is 
dominated by the saddle point $s^*$. Assuming that $1-s^* = O(1)$ as $V\to\infty$, the saddle point
is given by $\phi'(s^*)=0$, where
\be
\label{eq:phip_a}
  \phi'(s) = -\frac{1}{s}\bigl[\rho - \rho_1(\beta,s)\bigr],
\ee
where, from \eqref{eq:Nb},
\be
  \rho_1(\beta,s) = -\frac{s}{\beta^2}\, J'(s)  = \frac{1}{\beta^2}\, I(s),
\ee
with $I(s)$ given in \eqref{eq:Is}.
The saddle point is thus the solution of
\be
\label{eq:sp_a}
  \rho_1(\beta,s^*) = \rho.
\ee

Since $I(1) < \infty$, it follows that 
\be
  \rho_1(\beta,s) \leq \rho_1(\beta,1) \equiv \rho_{\rm c}(\beta) = \frac{1}{\beta^2}\, I(1).
\ee

As a consequence, the saddle point equation \eqref{eq:sp_a} does not have a solution for fixed
$\rho$ if $\beta > \beta_{\rm c}(\rho)$ with
\be
  \beta_{\rm c}(\rho): \qquad \rho_{\rm c}(\beta_{\rm c})= \rho,
\ee
or, alternatively, for fixed $\beta$ if $\rho > \rho_c(\beta)$.

To study this regime, we consider
\be
\label{eq:phib_a}
\begin{split}
  \ov{\phi}(s) &= -\rho\,\ln s + \frac{1}{V}\ln\, Z_\GCe(\beta,\alpha_0^{-1}s)
     \\
                    &= -\frac{1}{V}\ln(1-s) + \phi(s).
\end{split}                    
\ee

A simple calculation shows that
\be
  \ov{\phi}'(s) = -\frac{1}{s}\bigl[\rho - \ov{\rho}(\beta,s)\bigr],
\ee
where
\be
   \ov{\rho}(\beta,s) = \frac{1}{V}\ov{N}(\beta,V, s)
                               = \frac{s}{V}\frac{\partial}{\partial s} \ln \, Z_\GCe(\beta,\alpha_0^{-1}s).
\ee

As a consequence, in the thermodynamic limit, the leading contribution to the integral
\be
  Z_\Ce(\beta,V, N) = \frac{\alpha_0^N}{2\pi i} \oint_{\Gamma} 
    \frac{ds}{s} \, \E^{V\ov{\phi}(s)}
\ee
comes from the region close to the saddle point $s^*$ given by
\be
  \ov{\rho}(\beta,s^*) = \rho.
\ee

Substituting \eqref{eq:phib_a}, the saddle point equation becomes
\be
\label{eq:phibp_a}
  \ov{\phi}'(\beta,s^*) = \frac{1}{V}\frac{1}{1-s^*} + \phi'(s^*) =0
\ee
which, using \eqref{eq:phip_a}, gives
\be
\label{eq:sp_aa}
  \frac{1}{V}\frac{s^*}{1-s^*} = \rho - \rho_1(\beta,s^*).
\ee

From this, we see that, if $\rho > \rho_{\rm c}(\beta)$, then $1-s^*$ must be positive and, moreover,
$1-s^* = O(1/V)$ as $V\to\infty$. From \eqref{eq:sp_aa}, it follows that
\be
   s^* = 1 - \frac{1}{V}\frac{s^*}{\rho-\rho_1(\beta,s^*)},
\ee
which, for $V\to\infty$, gives
\be
\label{eq:s_star}
  s^* \sim 1 - \frac{1}{V}\frac{1}{\rho - \rho_{\rm c}(\beta)} + O(1/V^2),
  \qquad V\to\infty.
\ee

Since $s^* < 1$, in \eqref{eq:ZC_aa}, we can integrate over the circle 
$C^*: s=s^*\E^{i\theta}$;  thus, in the limit $V\to\infty$,
\be
\begin{split}
  Z_\Ce(\beta,V, N) = \frac{\alpha_0^N}{2\pi i} \oint_{C^*} 
    \frac{ds}{s} \, \frac{\E^{V\phi(s)}}{1-s}
    &\sim 
    \frac{\alpha_0^N}{2\pi i} \int_{s=s^*\E^{i\theta}, |\theta|<\eps} 
    \frac{ds}{s} \, \frac{\E^{V\phi(s)}}{1-s}
    \\
    &\sim
    \frac{\alpha_0^N}{2\pi i} \int_{s=s^* + iy, |y|<\eps} 
    \frac{ds}{s} \, \frac{\E^{V\phi(s)}}{1-s},
\end{split}
\ee
where $\eps$ is a small (arbitrary) parameter used to isolate the relevant region of integration.
Since $\eps$ is small, we can expand $\phi(s)$ as
\be
  \phi(s^*+iy) \sim \phi(s^*) + \phi'(s^*) iy - \frac{1}{2}\phi''(s^*)y^2 + O(y^3),
\ee
where, from \eqref{eq:phibp_a} and \eqref{eq:s_star},
\be
  \phi'(s^*) = -\frac{1}{\eta}, \qquad \eta^{-1} = \rho - \rho_{\rm c}(\beta),
\ee
while
\be
  \phi''(s^*) = -\frac{1}{s^*}\phi'(s^*)  + \frac{1}{s}\frac{\partial}{\partial s}\rho_1(\beta,s)\Bigr|_{s=s^*}  > 0.
\ee

Substituting, we then have, as $V\to\infty$,
\be
  Z_\Ce(\beta,V, N)  \sim
     \frac{\alpha_0^N}{2\pi } \int_{-\infty}^{+\infty}
     \frac{dy}{1 - \eta/V +iy} \, \frac{\E^{V\bigl[\phi(s^*) - iy/\eta -\frac{1}{2}\phi'' y^2
     \bigr]}}{\eta/V-iy} \bigl[1 + O(Vy^3)\bigr].
\ee

The proposed approach follows that described by Dingle in Ref. \cite{dingle1973asymptotic}, where he 
performs an additional change in variable, 
namely, takes $y=t/\sqrt{V\phi''}-i\eta/V$ and expands the term $O(Vy^3)$ and
$(1-\eta/V+iy)^{-1}$ in powers of $t$. This is equivalent to taking
$s= 1 + it/\sqrt{V\phi''}$ as the integration path.
Since we are only interested in the leading term of $\ln Z_\Ce(\beta,V, N)/V$ 
as $V\to\infty$, this step is not necessary, and we can just expand in powers of $V^{-1}$. 
Thus, we have
\be
  Z_\Ce(\beta,V, N)  \sim
     \frac{\alpha_0^N \E^{V\phi(s^*)}}{-2\pi i } \int_{-\infty}^{+\infty}
     dy\, y^{-1} \, \E^{ - iyV/\eta -\frac{V}{2}\phi'' y^2} + \ldots,
\ee
where the dots denote sub-leading terms as $V\to\infty$. 
By rescaling $y \to y/\sqrt{V\phi''}$, we finally arrive at
\be
  Z_\Ce(\beta,V, N)  \sim
     \frac{\alpha_0^N \E^{V\phi(s^*)}}{-2\pi i } \int_{-\infty}^{+\infty}
     dy\, y^{-1} \, \E^{ iya -\frac{1}{2} y^2} + \ldots,
\ee
where $a= -\sqrt{V/\phi''}/\eta$. 
The integral can be expressed in terms of the the parabolic cylinder function $D_{-1}(a)$:
\be
  \int_{\-\infty}^{+\infty} \frac{dy}{y} \E^{-\frac{1}{2}y^2 +iay} = -i\sqrt{2\pi}\,\E^{a^2/4} D_{-1}(a),
\ee
so that
\be
  Z_\Ce(\beta,V,N)  \sim
     \frac{\alpha_0^N \E^{V\phi(s^*)}}{\sqrt{2\pi} } \,\E^{a^2/4} D_{-1}(a)
     + \ldots.
\ee

In the limit $V\to\infty$, the term $a$ is large and negative, $D_{-1}(a)\sim \E^{-a^2/4} $, and hence
\be
  Z_\Ce(\beta,V, N)  \sim
     \frac{\alpha_0^N \E^{V\phi(s^*)}}{\sqrt{2\pi} }
     + \ldots,  
     \qquad V\to\infty.
\ee

Finally,
\be
\begin{split}
  \phi(s^*) &= -\rho\ln s^* -\frac{1}{\beta^2}J(s^*)
  \\
              & \sim -\rho\ln(1-\eta/V) - \frac{1}{\beta^2}J(1-\eta/V)
              \\
              &\sim \rho\eta/V - \frac{1}{\beta^2}J(1) + \frac{1}{\beta^2}J'(1)\eta/V + \ldots
\end{split}               
\ee

But, $J'(1)/\beta^2 = -\rho_{\rm c}(\beta)$; thus,
\be
\begin{split}
  \phi(s^*) &\sim [\rho - \rho_{\rm c}(\beta) ]\eta/V - \frac{1}{\beta^2}J(1)  + \ldots
                \\
                &\sim 1/V - \frac{1}{\beta^2}J(1)  + \ldots.
\end{split}               
\ee

In conclusion, the leading term $\ln Z_\Ce(\beta,V,N)/V$ as $V\to\infty$ is
\be
\label{eq:Z_clead}
  Z_\Ce(\beta,V,N)  \sim \alpha_0^N\, \E^{ -\frac{V}{\beta^2}\,J(1)}
     \qquad V\to\infty.  
\ee

We can now go back to \eqref{eq:KNz-GC_1}, which, using \eqref{eq:Z_clead}, becomes
\be
  K(N|z) \sim \frac{(z\alpha_0)^{N}}{Z_\GCe(\beta,V, z)}\,  \E^{ -\frac{V}{\beta^2}\,J(1)},
  \qquad V\to\infty.
\ee

Now, recalling that the fugacity $z$ is related to $\langle N\rangle_\GCe$ via \eqref{eq:rho_av_z},
\be
  (z\alpha_0)^{N} = \E^{N\ln(z\alpha_0)} \sim \E^{N\ln(1 - \ov{\eta}/V)} 
          \sim \E^{-\rho\, \ov{\eta}},
\ee
because $\alpha_0 z = 1 - \ov{\eta}/V$ with $\eta^{-1} = \ov{\rho} - \rho_{\rm c}(\beta)$.
Similarly,
\be
\begin{split}
  \E^{ -\frac{V}{\beta^2}\,J(1)} Z_\GCe(\beta,V,s)^{-1} &=
               (1-\alpha_0 z) \E^{-\frac{V}{\beta^2}\bigl[J(1) - J(z\alpha_0)\bigr]}
               \\
               &\sim
               \frac{\ov{\eta}}{V}\, \E^{-\frac{V}{\beta^2} J'(1)\,\ov{\eta}/V} 
               \sim  \frac{\ov{\eta}}{V}\, \E^{-\rho_{\rm c}(\beta)\,\ov{\eta}},
               \qquad V\to\infty.
\end{split}  
\ee

Thus, collecting all the terms we have, cfr. \eqref{eq:Kz_crt} and \eqref{eq:Krr},
\be
\begin{split}
  K(N|z) = K(\rho|\ov{\rho})\,\Delta\rho  &\sim \frac{\ov{\eta}}{V}\, \E^{-[\rho-\rho_{\rm c}]\,\ov{\eta}} 
  \\
             &\sim \Delta\rho\, \frac{\E^{-(\rho-\rho_{\rm c})/(\ov{\rho} - \rho_{\rm c})}}{\ov{\rho} - \rho_{\rm c}},
             \qquad \rho> \rho_{\rm c}(\beta),\ V\to\infty.
\end{split}             
\ee
where $\Delta\rho = 1/V$. 


\end{document}